\documentclass[manuscript,screen,nonacm]{acmart}

\AtBeginDocument{%
  }

\usepackage{tabularx}

\usepackage{longtable}
\usepackage{array}
\usepackage{soul} 

\usepackage{tcolorbox}
\tcbuselibrary{listings, breakable}
\newcolumntype{L}{>{\raggedright\arraybackslash}p{0.12\textwidth}}
\newcolumntype{M}{>{\raggedright\arraybackslash}p{0.29\textwidth}}
\newcolumntype{C}{>{\centering\arraybackslash}p{0.08\textwidth}}

\sethlcolor{yellow}
\begin{document}

\title{A Study on the Impact of Fault localization Granularity for Repository-Scale Code Repair Tasks}


\author{Joseph Townsend}
\email{joseph.townsend@fujitsu.com}
\orcid{<ORCID-ID>}
\affiliation{%
  \institution{Fujitsu Research of Europe}
  \city{London}
  \country{UK}
}
\authornote{Equal contribution to this research.}

\author{Chandresh Pravin}
\email{chandresh.pravin@fujitsu.com}
\orcid{<ORCID-ID>}
\affiliation{%
  \institution{Fujitsu Research of Europe}
  \city{London}
  \country{UK}
}
\authornotemark[1]

\author{Kwun Ho Ngan}
\email{kwun.hongan@fujitsu.com}
\orcid{<ORCID-ID>}
\affiliation{%
  \institution{Fujitsu Research of Europe}
  \city{London}
  \country{UK}
}
\authornotemark[1]

\author{Matthieu Parizy}
\email{parizy.matthieu@fujitsu.com}
\orcid{<ORCID-ID>}
\affiliation{%
  \institution{Fujitsu Research of Europe}
  \city{London}
  \country{UK}
}


\begin{abstract}
    Automatic program repair (APR) can be a difficult and demanding task, especially when it comes to resolving complex issues at a repository-level. Resolution often includes issue reproduction, fault localization, code repair, testing and validation. Issues of this scale can be commonly found in popular GitHub repositories or datasets that are derived from them such as SWE-Bench \cite{jimenez2024swebenchlanguagemodelsresolve}.

    Some repository-level approaches separate localization and repair into distinct phases. Where this is the case, the fault localization (FL) approaches vary in terms of the granularity of localization. For example, some of these frameworks will identify specific methods or functions responsible for an issue \cite{chen2024coder, wadhwa2024masai}, others will narrow down to specific code lines or segments of lines \cite{liu2024marscode, xia2025demystifying, ma2025swe}. Some even localize to entire files before passing this information on to the repair agent \cite{tao2024magis, wei2025swe}. However, there does not appear to have been much effort towards an empirical study of what level of granularity is most beneficial for the repair tasks.
        
    Where the impact of granularity is explored to some degree for smaller datasets, not all isolate this issue from the separate question of localization accuracy by testing code repair under the assumption of perfect fault localization. To the best of the authors' knowledge, no repository-scale studies have explicitly investigated granularity under this assumption, nor conducted a systematic empirical comparison of granularity levels in isolation.
        
    We propose a framework for performing such tests by modifying the localization phase of the \emph{Agentless} framework \citep{xia2025demystifying} to retrieve ground-truth localization data and include this as context in the prompt fed to the repair phase. 
    We show that under this configuration and as a generalization over the SWE-Bench-Mini dataset, function-level granularity yields the highest repair rate against line-level and file-level. However, a deeper dive suggests that the ideal granularity may in fact be task dependent. 
    
    This study is not intended to improve on the state-of-the-art for SWE-Bench-Mini or any other dataset, nor do we intend for results to be compared against any complete agentic frameworks. Rather, we present a proof of concept for investigating how fault localization may impact automatic code repair in repository-scale scenarios. We present preliminary findings to this end and encourage further research into this relationship between the two phases. 
\end{abstract}




\maketitle



\section{Introduction}
\label{sec:intro}

Over the past decade, automated program repair (APR) has advanced significantly from its origins, fuelled by more capable program‑analysis tools, the emergence of large language models, and rapid progress across machine learning as a whole. 

Great efforts are being made to perform repository-level code repair with the help of LLM-based agents, especially in popular, highly competitive challenges such as SWE-Bench \cite{jimenez2024swebenchlanguagemodelsresolve}. This scale of repair is highly important because unlike many of the smaller datasets used to evaluate code repair methods \cite{austin2021program, chen2021evaluating}, repository-level repair tasks are derived from genuine code issues encountered by software developers in their daily tasks. The scale of issues represented by smaller datasets tends to be limited to single functions or files, and may only require correction of single code lines in order to be resolved.

Some of these repository-level approaches separate localization and repair into distinct phases by providing different instructions in the prompt or even by using different agents. Where this is the case, the fault localization approaches vary in terms of the granularity of localization. For example, some of these frameworks will identify specific methods or functions responsible for an issue \cite{chen2024coder, wadhwa2024masai}, others will narrow down to specific code lines or segments of lines \cite{liu2024marscode, xia2025demystifying, ma2025swe}. Some even localize to entire files before passing this information on to the repair agent \cite{tao2024magis, wei2025swe}. However, there does not appear to have been much effort towards an empirical study of what level of granularity is most beneficial for the repair tasks.

Many fault localization approaches merit themselves on high localization scores, but in repair tasks, a high or even perfect localization score has no value if it does not influence the repair agent in the generation of a patch which resolves the initial issue. Thus, we argue that before investigating new means of fault localization for the purposes of program repair, it is prudent to first explore suitable means of presenting perfect fault localization data to the repair agent, which is possible using ground truths provided in fault localization datasets. In other words, if there is no format under which the ground truths of localization data influence the repair agent in generating patches more successfully, then there is no value in attempting to model fault localization for that dataset until the shortfalls of the repair agent have been addressed. To the authors' knowledge, the literature does not explore this in detail either.

While code-line level fault localization is uncommon in repository-level tasks, it is more commonly applied to smaller datasets that don't capture the realism of repository-scale tasks derived from genuine source code. Where developers do intend to upscale the scope of fine-grained methods from some of these smaller datasets to larger, more realistic contexts, we encourage them to first consider experiments such as these "perfect localization" tests. We are not trying to argue that there is no place for fine-grained localization, only that there is a case for more methodical exploration of its application (or that of any granularity of localization) to real-world tasks before further development.

Section \ref{sec:background} provides a review of background material, observing that despite a variety of fault-localization guided APR methods being proposed for repository-scale tasks, very few test the perfect localization hypothesis to back-up the validity of treating these two tasks as separate phases, and also that very little effort is made to confirm a suitable level of granularity. Section \ref{sec:method} presents a proof of concept for performing such tests by modifying the localization phase of the \emph{Agentless} framework \citep{xia2025demystifying} to retrieve ground-truth localization data and including this as context in the prompt fed to the repair phase (which in our case is Qwen3-Coder-30B~\cite{qwen3technicalreport}). We also bypass Agentless' patch validation phase to eliminate the influence of the re-ranking process on the final resolve rate, thus isolating the impact of localization granularity even further. Section \ref{sec:results} presents the findings, namely that under this configuration at least function-level granularity yields the highest resolve rate for the SWE-Bench-Mini subset of the SWE-Bench Verified benchmark overall, but a deeper dive suggests that the ideal granularity may in fact be task-dependent. Section \ref{sec:Limitations} highlights the limitations and the boundaries which are critical in the automatic program repair (APR) but will not be assessed in this work. Section \ref{sec:future_directions} proposes further variations of our study to expand the scope of these experiments and observe whether or not the same findings hold. Section \ref{sec:conclusions} concludes.

\section{Background}
\label{sec:background}

Automated program repair (APR) pipelines typically involve multiple stages, including issue reproduction, fault localization (FL), patch generation, and validation. Fault localization and patch generation are two critical components that are commonly delegated to separate agents or distinct passes in large-scale repository-level repair~\cite{wang2023rap,hasen2025Autostructor,yang2025patch}. The recent emergence of LLM-based approaches have further supplemented this design using multi-agent and multi-phase pipelines~\cite{xia2025demystifying,prenner2022can}. These approaches can be broadly classified into two frameworks: procedural frameworks that decompose the code repair task into a fixed sequence of discrete steps~\cite{yang2025survey}, and agentic frameworks that use LLMs to plan and execute code repair steps with access to tools such as code search and test execution~\cite{bouzenia2024repairagent,yang2024swe}. Although alternative approaches such as fine‑tuning and generate‑and‑validate strategies exist, our discussion centers on fault localization and patch generation as two core components that appear across a broad range of APR systems, rather than suggesting a rigid two‑stage pipeline~\cite{monperrus2018living}. The primary standard by which the field measures progress is SWE-Bench~\cite{jimenez2024swebenchlanguagemodelsresolve}, which has become the benchmark for evaluating LLM-based APR systems on a realistic repository-level scale~\cite{yang2024swebench}. Despite this, the granularity and representation of localization information remains an open question~\cite{cao2023code}, with little empirical evidence presented for these design choices~\cite{liu2019You,zhang2024systematic,yang2025survey}.

Numerous APR systems illustrate the two-stage fault-localization and patch-generation pipeline~\cite{yang2025survey,Monperrus2018automatic}. Agentless~\citep{xia2025demystifying} follows a hierarchical localization process, first narrowing to suspicious files, then to relevant classes or functions, and finally to fine-grained edit locations before passing these to its repair phase. AutoCodeRover~\cite{zhang2024autocoderover} takes a software-engineering-oriented approach, using structured code search over AST representations at the class and method level, supplemented by method-level spectrum-based fault localization where a test suite is available. MASAI~\cite{Masi2024Arora} decomposes the task across multiple specialised sub-agents, with its localization sub-agent tasked specifically with identifying files to edit via multi-step reasoning over the issue description. By contrast SWE-Agent~\cite{yang2024swe} adopts a more agentic approach where the fault localization is not a discrete step but emerges from an iterative interaction through a command-line interface. Whilst all of these methods approach the bug repair task through some form of fault localization prior to patch generation, the granularity at which the localization signal is first derived and then propagated to the patch generation phase differs across frameworks~\cite{yang2025survey,Zhang2023surveylearning}, from file-level in some instances, to function and edit locations in others. Importantly, in no case is this choice empirically justified. This inconsistency suggests that granularity of localization is treated as an implementation detail rather than a principled design decision, and motivates a more systematic investigation.

The variation in localization granularity across APR systems is not only a question of localization precision, rather, a fundamental question on \textit{what} information the model requires to generate a patch successfully. Coarser granularity, such as file-level localization, provides the repair agent with broad surrounding context at the cost of specificity, requiring the model to identify the precise edit location independently~\cite{yen2025helmet,rando2025longcodebench}. Finer granularity, such as line-level localization, provides high specificity but may omit contextually relevant code (e.g., call sites, dependency definitions, or test constraints) presented in way that the repair agent requires to generate a correct patch~\cite{park2025emulating,wang2025coderag,zhang2025coderag}. The granularity of localization therefore mediates a trade-off between specificity and context that has direct implications for repair agent performance, independent of how accurately the localization was derived.

This trade-off has received substantial attention in the fault localization literature but almost exclusively as a property of the localization step itself, evaluated in terms of precision and recall at file, method, or line level~\cite{sarhan2022SBFLsurvey}. 
In contrast, we consider localization granularity not only as the spatial scope of the identified region (e.g., file, function, or line), but also as how that region is presented to the repair agent, in terms of the amount and structure of code context provided.
The question then of how these factors affect the downstream repair agent, rather than the localization phase alone, has been largely overlooked previously~\cite{Zhang2023surveylearning}.
To the extent that granularity is not discussed in APR systems at all, it is treated as a fixed architectural choice rather than a variable with measurable downstream impact on patch generation success. This distinction motivates a direct empirical investigation of how different granularities of ground-truth localization information influence repair outcomes, independently of any localization method.

One possible but under-explored explanation for this inconsistency is that LLMs used for code repair exhibit an implicit bias toward the granularity of code at which they were pre-trained or instruction-tuned~\cite{fried2023incoder,li2025exploratory}. Open code generation LLMs show that, where this information is available, are typically trained on file and repository-level granularities with extended sequences and infilling tasks.  DeepSeek‑Coder~\cite{guo2024deepseekcoder,deepseekai2024deepseekcoderv2}, for instance, uses repo‑level pretraining, whilst StarCoder2~\cite{lozhkov2024starcoder2stackv2} is trained using repository-level context with file‑level fill-in-the-middle transformations. InCoder~\cite{fried2023incoder} is trained to generate full code files by means of masking file-level code. Qwen3-Coder-Next~\cite{qwen3codernexttechnicalreport}, on the other hand, shows examples of function/class level bug synthesis to scale up their agentic training procedure, which follows from the Qwen2.5-Coder~\cite{huiqwen25coder2025} using file-and-repo-level training. Whilst evidence from code generation settings suggests that fine-tuning data granularity influences downstream task performance~\cite{li2025exploratory}, whether this effect holds for large-scale repository-level APR has yet to be tested. Our work isolates this question by removing the localization error as a source of bias, providing a direct empirical evaluation of the effects of granularity on patch‑generation success, independent of any localization method.  


    
    
    
    
    


\section{Method}
\label{sec:method}

Motivated by a lack of empirical justification for choice of granularity in repository-scale APR, we describe our method for performing such an evaluation at three levels: line-level, function-level and file-level. Furthermore, key to our approach is to adopt the perfect localization assumption so that analysis of the impact of each granularity can be concentrated purely on the repair phase.

\subsection{Dataset}
\label{subsec:method_dataset}

We evaluate on the SWE-Bench Mini dataset\footnote{https://huggingface.co/datasets/MariusHobbhahn/swe-bench-verified-mini}, which is a subset of 50 instances from the larger \emph{SWE-Bench Verified}\footnote{https://www.swebench.com/verified.html} set, containing 500 instances each derived from human-verified GitHub issue resolutions across 12 popular, open-source python repositories. SWE-Bench Mini's 50 instances are taken from only 2 repositories: \emph{django}\footnote{https://github.com/django/django} and \emph{sphinx-doc}\footnote{https://github.com/sphinx-doc/sphinx}. Each instance contains both the broken code before repair and the solution code (i.e. the pull request). Both datasets are vetted by humans, so the ground truth patches which form the basis of the localized context in our perfect localization assumption can be assumed relevant to the issues to be solved. 

\subsection{Architecture}
\label{subsec:method_dataset}

\begin{figure}
    \centering
    \includegraphics[width=0.8\linewidth]{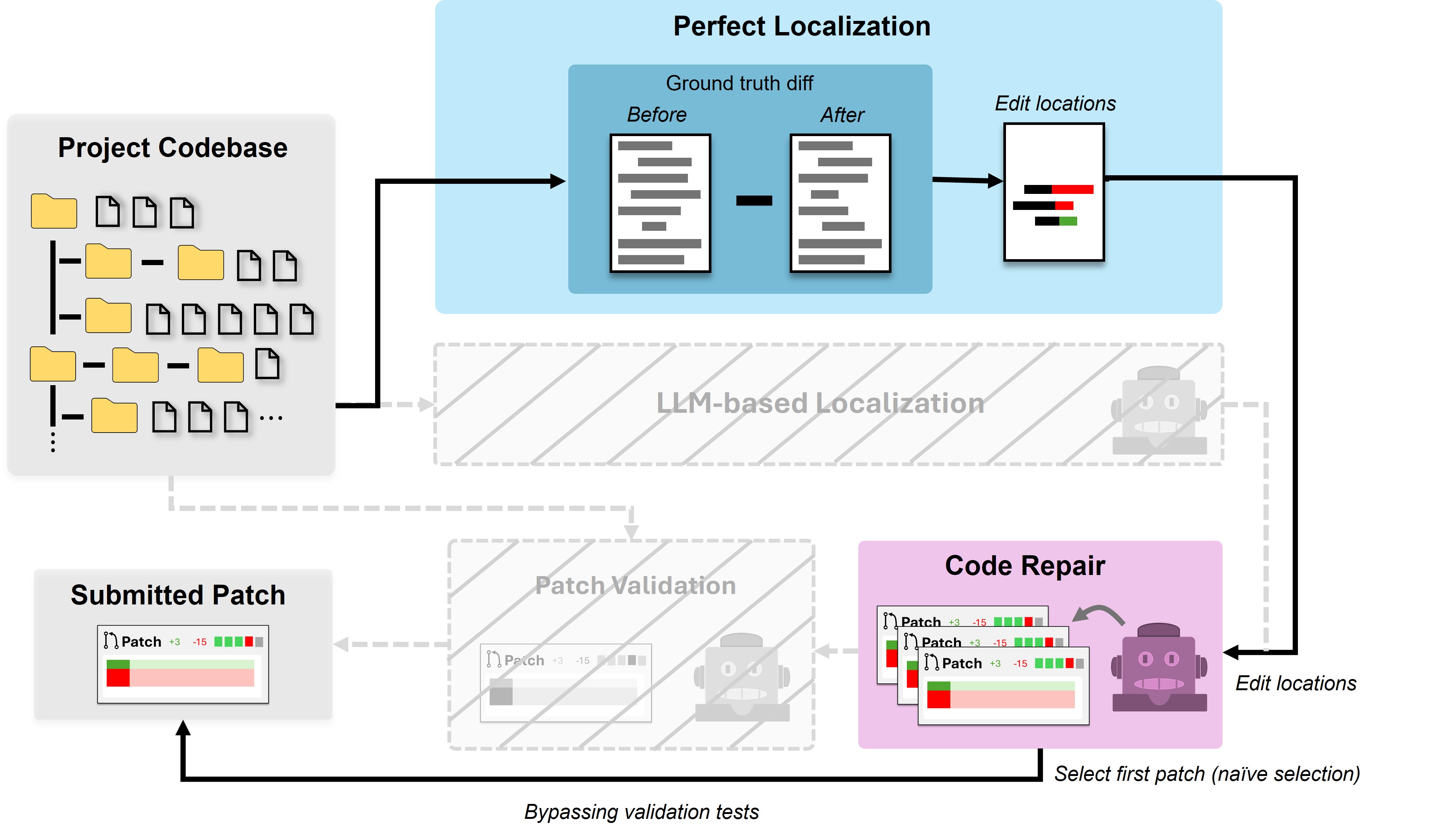}
    \caption{A modified version of the Agentless framework \citep{xia2025demystifying} was designed to conduct the experiments. The file localization and patch validation phase from the original work were bypassed such that a replacement script derives perfect localization data from the ground truth patches to produce the raw generated patches for SWEBench evaluation in order to concentrate the study on the effect of localization granularity.}
    \label{fig:placeholder}
\end{figure}

We chose Agentless \citep{xia2025demystifying} to orchestrate the patch generation pipeline as a simple architecture that enabled us to focus on the changes we were interested in. It is sequential in nature, enabling the isolation of individual phases and avoiding the complexity of agentic interactions found in environments such as OpenHands \cite{wang2024openhands} and SWE-Agent \cite{yang2024swe}.

We bypass Agentless' fault localization phase to provide ground truth localizations to the repair phase, thus implementing the "perfect localization" assumption necessary to investigate the effect of localization on code repair. How we obtain the ground truth localizations and feed these into the repair phase are covered in section \ref{subsec:method_localization}. For the code repair phase, we employ Qwen3-Coder-30B \cite{qwen3technicalreport} as the coding language model, deployed via vLLM v0.15.1 \cite{kwon2023efficient}. Finally, we also bypass Agentless' patch validation phase, selecting the first patch generated by default. The motivation for this to isolate the influence of the localization phase independently of any re-ranking. 


\subsection{Localization}
\label{subsec:method_localization}

We adopt the "perfect localization" assumption, in which fault localization data is taken from the ground truth and injected into the repair prompt. This assumption has been criticised on the grounds that it is unrealistic \cite{campos2025empirical}. While we do agree with this point, we believe the assumption is necessary when attempting to isolate and inspect other aspects of an APR method's performance. In our special case of attempting to identify an appropriate level of localization granularity for an APR agent, it would be a distraction if the accuracy of the localization method for each granularity were called into question. Thus, by adopting the perfect localization assumption, we isolate the granularity as the only variable. For an evaluation of FL and APR in concert, we would absolutely encourage the use of genuine localization.

\begin{figure}
    \centering
    \includegraphics[width=\linewidth]{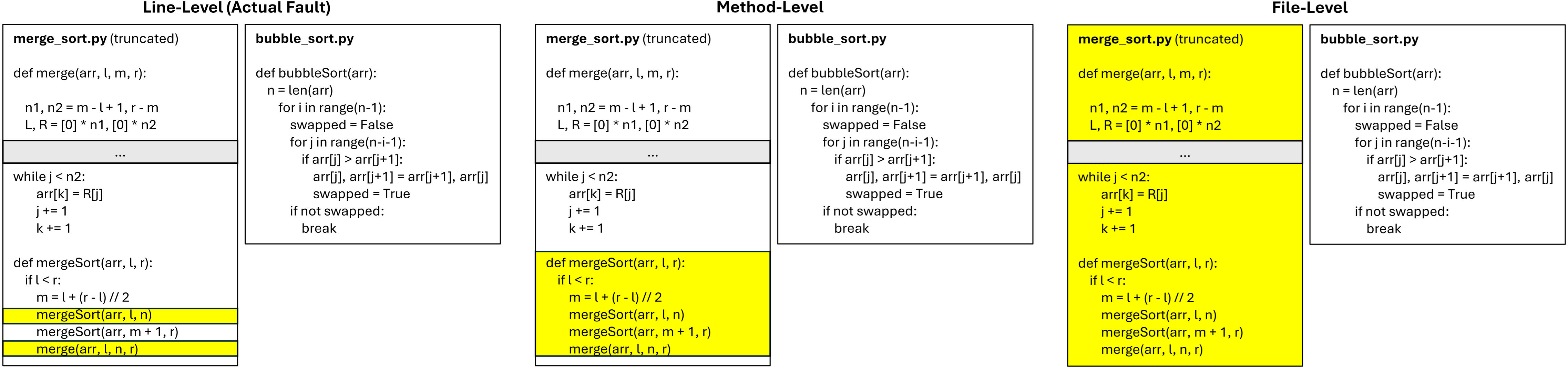}
    \caption{Fault localization at various levels of granularity, demonstrated in a hypothetical repository containing two sorting methods. The actual fault is in the \emph{mergeSort} method of \emph{merge\_sort.py}, where \emph{n} should be replaced with \emph{m}. At line-level granularity, only these two code lines are provided as context for the repair phase, indicated by the yellow highlight. At function-level, the entire function is provided. And at file-level, the entire \emph{merge\_sort.py} is provided.}
    \label{fig:granularity_comparison}
\end{figure}

We obtain the ground truth localizations based on the code diffs generated for an issue's code base before and after (i.e. the ground truth repair solution) repair in the dataset. For line-level, this is therefore each exact line that is changed in the diff. For function-level, we use any function or class that contains lines from the line-level analysis, and for file-level we use any files that contain modified functions or classes. Note also that while we acknowledge the difference between a method and a function, for reporting purposes we refer to both as `function' and treat them and any classes that contain them likewise in the experiments. Figure \ref{fig:granularity_comparison} provides examples of how the three degrees of localization would manifest for two erratic lines in a hypothetical repository of two files.

Given a set of fault localizations, there is also the question of how best to inject these into the prompt. Since it is known that LLMs struggle with counting \cite{yang2024number, zhang2024counting, xu2025llm}, we did not consider the option of simply listing line numbers for the repair phase to address in the prompt. Instead we included the actual code lines/functions (plus the 10 lines before and after, for context) as code snippets. The prompt templates may be found in the appendix.

\subsection{Evaluation}
\label{subsec:method_evaluation}

For each of the 3 granularities, we evaluate the pipeline 10 times for each instance, recording whether or not the corresponding issue resolves or if the execution even completes. For each trial, we count the total number of resolutions, failures to resolve, and completions, and calculate the resolve rate as a percentage of the total 50 instances. For each granularity, we average all of these metrics across all 10 trials.

\section{Results}
\label{sec:results}

We consider the results from two perspectives: generalized across all 50 tasks in the dataset, and from the perspective of task difficulty.

\subsection{Generalized}
\label{subsec:results_general}

\begin{figure}
    \centering
    \includegraphics[width=0.75\linewidth]{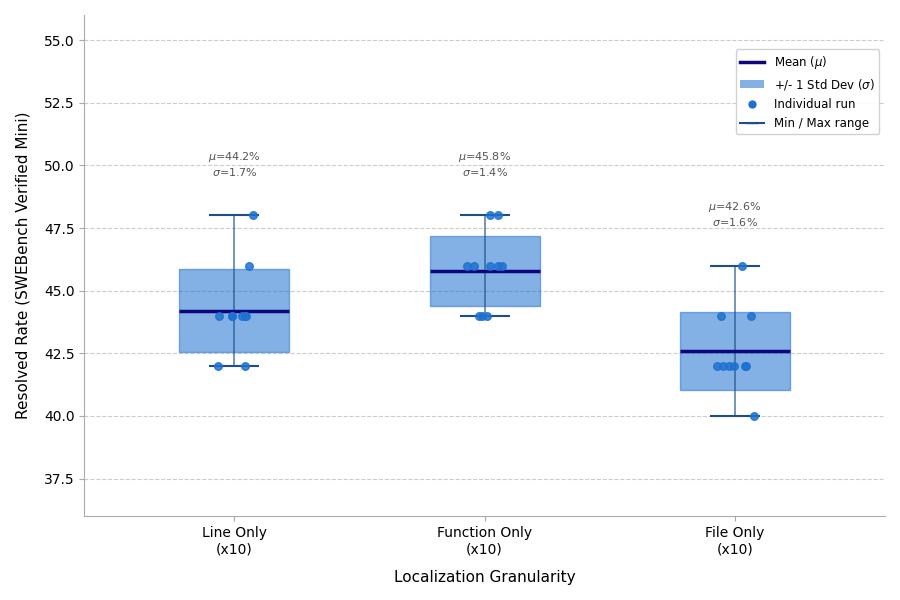}
    \caption{Box plot of resolve rate achieved for different granularities of perfect fault localization as reported in table \ref{tab:results}}
    \label{fig:results}
\end{figure}


\begin{table}[t]
    \centering
    \caption{Resolved rates and corresponding standard deviation ($\sigma$) for code repair using Qwen3-Coder under the Agentless framework on the SWE-Bench-Mini dataset. Repair prompts are derived from ground-truth fault localization at different granularities. Results are averaged over 10 trials.}
    \label{tab:results}
        \begin{tabular}{lcccccc}
        \toprule
        \textbf{Granularity} & \textbf{Resolved} & \textbf{Unresolved} & \textbf{Completed} & \textbf{Resolved Rate (\%)} & \textbf{$\sigma$ (Rate)} \\
        \midrule
        Line     & 21.8 & 26.8 & 48.6 & 43.6 & 2.6 \\
        Function & 22.8 & 27.2 & 50.0 & 45.6 & 1.5 \\
        File     & 21.3 & 27.5 & 48.8 & 42.6 & 1.6 \\
        \bottomrule
        \end{tabular}
\end{table}

The results of these experiments are shown in table \ref{tab:results} and figure \ref{fig:results}. A Friedman test on the results yields $\chi^2=12.8$, $\rho=0.0017$, indicating statistical significance. Resolved rate is the highest for the function-level granularity (i.e. for functions, methods and classes). Function-level also boasts the lowest standard deviation of the three granularities, indicating that it is the most stable over repeated runs. Given that this is the level of granularity at which Qwen3-Coder is trained, this is perhaps to be expected, but it would be worth verifying in future work whether or not this holds true for other models. In other words, is there empirical evidence that coding LLMs perform best at the level of granularity on which they were trained? 

Resolved rate is lowest for file-level granularity. Line-level yields the lowest standard deviation across 10 trials, indicating that its performance is the least stable.

Note also that the best resolve rate we achieve is below the 50\% mark, over 20\% behind the state-of-the-art achieved by \emph{Claude Sonnet 4.5}\footnote{https://hal.cs.princeton.edu/swebench\_verified\_mini, checked on 31st March 2026.}. This indicates a very important point: that even perfect fault localization does not necessarily yield good code repair.

\subsection{Split by difficulty}
\label{subsec:results_difficulty}

\begin{figure}
    \centering
    \includegraphics[width=0.75\linewidth]{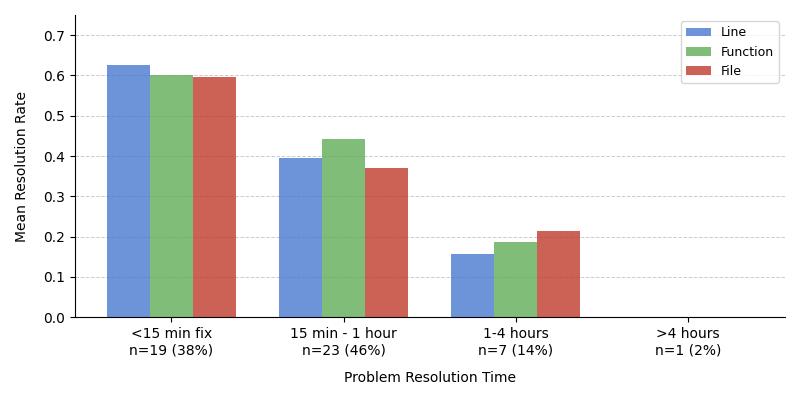}
    \caption{Mean resolution rate by problem difficulty tier and localization granularity. Difficulty is inferred from human annotator resolution time as provided in the SWE-Bench Verified metadata. Results are averaged over 10 trials per granularity.}
    \label{fig:difficulty}
\end{figure}

Furthermore, figure~\ref{fig:difficulty} stratifies the resolve rate by instance difficulty, as labelled in the SWE-Bench Verified metadata based on the time taken by a human annotator to resolve each issue. Whilst function-level granularity dominates on average, the relative ordering across granularities shifts with difficulty tier, suggesting that the optimal level of localization granularity is not fixed but instance-dependent. This finding motivates further investigation into whether difficulty-aware or adaptive granularity selection could improve repair outcomes beyond what any single fixed granularity achieves.

\section{Limitations}
\label{sec:Limitations}

This study is not intended to achieve the state-of-the-art code repair performance on SWE-Bench-Mini or similar benchmarks. Nor is our approach intended to represent a full agentic repair system as presented in other work. Instead, we focus specifically on understanding how fault localization, and in turn the generated context, affects automatic program repair, and present our findings as a preliminary proof of concept in this research direction.

Based on this objective, we deliberately simplified the APR pipeline and excluded factors such as test-time scaling (TTS), prompt engineering, and action planning as implemented in conventional coding agents. Although these components may improve resolution performance, these factors would also introduce additional confounding variability making it hard to isolate the contribution purely from fault localization granularity.

This study is also limited by the definition of "perfect" fault localization. We approximate this by using the set of code changes appearing in the associated pull request. This however may ignore code that may actually be relevant to the fault as context but does not require modification for issue resolution. Determining which code is truly relevant to an issue remains an open research question and lies beyond the scope of this paper.



\section{Future Directions}
\label{sec:future_directions}

Perhaps the most logical next step is to observe whether these observations extend to the larger SWE-Bench Verified from which the mini dataset is derived, especially since SWE-Bench mini contains instances from only 2 public repositories out of the total 12 used in SWE-Bench Verified. That said, recently significant weaknesses of SWE-Bench Verified as a means of evaluation have been identified by its curators\footnote{https://openai.com/index/why-we-no-longer-evaluate-swe-bench-verified/}, arguing that SWE-Bench Pro\footnote{https://labs.scale.com/leaderboard/swe\_bench\_pro\_public} may now be a more suitable benchmark. Regardless of which is more suitable, the real concern as that the experiments in this paper should be repeated on even larger benchmarks.

Similar evaluation with other coder models would also be welcome. In particular this would be useful to evaluate the hypothesis that suitable granularity for repair aligns with the granularity on which the models were trained.

Also worth exploring would be combinations of some of the granularities discussed, for example combining function-level and file-level by extracting code blocks at the function level and commenting the code lines identified in the original line-level ground truth. Another variable related to how localization information is presented is the context window, and the impact of this on performance would also be interesting to explore.

More generally, a wider exploration of the relationship between a prompt template and resolve rate would also be beneficial, but also very challenging. There are many factors other than localization granularity to consider such as whether or not to include error traces, issue descriptions, etc. The space of possible prompt templates is astronomically large. If we are going to continue injecting knowledge into code repair prompts, a more systematic approach for searching the space of possible templates is needed, and as much as possible the question of what template is appropriate for the APR must be isolated from the question of how reliable the injected knowledge is (for example, by using ground truths when that knowledge is the localized faults).

Finally, future work should also incorporate the fault localization module from the original Agentless framework \cite{xia2025demystifying} into the experiments, in order to compare Agentless' localization quality against our proxy for perfect localization and to analyze any additional contextual information it identifies.

\section{Conclusions}
\label{sec:conclusions}

We have shown that for the SWE-Bench-Mini benchmark, when repaired using Qwen3-Coder in the Agentless framework \cite{xia2025demystifying},  function-level granularity fault localization  yields the highest overall resolve rate when that localization data is input as context to the repair prompt. This level of granularity also offers the best stability, as over 10 trials the variation of resolve rate was the smallest. In comparison, file-level localization yielded the lowest resolve rate and line-level the greatest variation. That being said, these findings are a generalization over the dataset. A deeper dive into how results vary with respect to difficulty suggested that the impact of fault localization is more complex, and that the ideal granularity may in fact depend on the task itself.

The next logical step for future work is to observe whether these same conclusions hold true for other, larger repository-level datasets, for other coding models, and for other variations on prompt template. 

We are the first to conduct a fault granularity analysis under the perfect localization assumption on SWE-Bench-Mini, or on any repository-scale automatic code repair task. This was not an attempt to improve on the state-of-the-art or compare against other architectures. Rather, we hope our experiments encourage others in the field to exercise caution when selecting an appropriate degree of granularity when localizing faults for injection into code repair prompts, and recommend preliminary tests under the perfect localization assumption to this end.




\section*{Acknowledgements}

This work would not have been possible without valuable insight from Yuheng Huang and Ao Peng from the University of Tokyo. The former's work on fault localization \cite{huang2025risk} inspired and informed our own venture into the field, and discussions with both since then have influenced the ideas covered in this paper. The authors would also like to thank their colleagues Shanshan Yu and Satoshi Munakata for facilitating these discussions between research teams.

\bibliographystyle{ACM-Reference-Format}
\bibliography{references}

\appendix

\section{Appendix}

\subsection{Prompt Template}
Source file context constructed from fault localization output; granularity varies from file-level (entire file) to function-level (relevant functions with context window) to line-level (specific line intervals with context window). The prompt used for by the repair agent is as follows:

\begin{lstlisting}[breaklines=true, basicstyle=\small\ttfamily]
We are currently solving the following issue within our repository. Here is the issue text:
--- BEGIN ISSUE ---
{<@problem_statement@>}
--- END ISSUE ---

Below are some code segments, each from a relevant file. One or more of these files may contain bugs.
--- BEGIN FILE ---
```
{<@content@>}
```
--- END FILE ---

Generate *SEARCH/REPLACE* edits to fix the issue. DO NOT include any explanations or analysis. ONLY output the code block.

Every *SEARCH/REPLACE* edit must use this format:
1. The file path
2. The start of search block: <<<<<<< SEARCH
3. A contiguous chunk of lines to search for in the existing source code
4. The dividing line: =======
5. The lines to replace into the source code
6. The end of the replace block: >>>>>>> REPLACE

EXAMPLE OUTPUT (this is the ONLY acceptable format):
```python
### mathweb/flask/app.py
<<<<<<< SEARCH
from flask import Flask
=======
import math
from flask import Flask
>>>>>>> REPLACE
```
REQUIREMENTS:
- NO text before or after the ```python...``` code block
- PROPER INDENTATION is required
- Start your response directly with ```python
\end{lstlisting}

\noindent\textit{Note: Terms in \textbf{\textcolor{blue}{\{braces\}}} denote variable placeholders substituted at inference time.}

\vspace{0.5em}
\noindent
\begin{tabular}{@{}ll@{}}
    \toprule
    \textbf{Placeholder} & \textbf{Description} \\
    \midrule
    \texttt{\{problem\_statement\}} & The GitHub issue text provided as task input \\
    \texttt{\{content\}} & Source file contents provided for repair \\
    \bottomrule
\end{tabular}
\vspace{1em}

\newpage
\subsection{Line-level Granularity Content Example}

The \texttt{\{problem\_statement\}} and \texttt{\{content\}} placeholders at line-level granularity for instance \lstinline{django__django-11848} are shown below. Line-level localization identifies the modified line ranges, but we expand these into small code snippets, because LLMs require some form of surrounding context to generate valid edits.

\begin{lstlisting}[breaklines=true, basicstyle=\small\ttfamily]
<@problem_statement@>:

django.utils.http.parse_http_date two digit year check is incorrect
Description
	 
		(last modified by Ad Timmering)
	 
RFC 850 does not mention this, but in RFC 7231 (and there's something similar in RFC 2822), there's the following quote:
Recipients of a timestamp value in rfc850-date format, which uses a
two-digit year, MUST interpret a timestamp that appears to be more
than 50 years in the future as representing the most recent year in
the past that had the same last two digits.
Current logic is hard coded to consider 0-69 to be in 2000-2069, and 70-99 to be 1970-1999, instead of comparing versus the current year.

\end{lstlisting}

\begin{lstlisting}[breaklines=true, basicstyle=\small\ttfamily]
<@content@>:

### django/utils/http.py
...
    """
    # email.utils.parsedate() does the job for RFC1123 dates; unfortunately
    # RFC7231 makes it mandatory to support RFC850 dates too. So we roll
    # our own RFC-compliant parsing.
    for regex in RFC1123_DATE, RFC850_DATE, ASCTIME_DATE:
        m = regex.match(date)
        if m is not None:
            break
    else:
        raise ValueError("%r is not in a valid HTTP date format" % date)
    try:
        year = int(m.group('year'))
        if year < 100:
            if year < 70:
                year += 2000
            else:
                year += 1900
        month = MONTHS.index(m.group('mon').lower()) + 1
        day = int(m.group('day'))
        hour = int(m.group('hour'))
        min = int(m.group('min'))
        sec = int(m.group('sec'))
        result = datetime.datetime(year, month, day, hour, min, sec)
        return calendar.timegm(result.utctimetuple())
    except Exception as exc:
        raise ValueError("%r is not a valid date" % date) from exc


def parse_http_date_safe(date):
    """
    Same as parse_http_date, but return None if the input is invalid.
    """
    try:
        return parse_http_date(date)
...
\end{lstlisting}

\newpage
\subsection{Function-level Granularity Content Example}

The \texttt{\{problem\_statement\}} and \texttt{\{content\}} placeholders at function-level granularity for instance \lstinline{django__django-11848} are shown below.

\begin{lstlisting}[breaklines=true, basicstyle=\small\ttfamily]
<@problem_statement@>:

django.utils.http.parse_http_date two digit year check is incorrect
Description
	 
		(last modified by Ad Timmering)
	 
RFC 850 does not mention this, but in RFC 7231 (and there's something similar in RFC 2822), there's the following quote:
Recipients of a timestamp value in rfc850-date format, which uses a
two-digit year, MUST interpret a timestamp that appears to be more
than 50 years in the future as representing the most recent year in
the past that had the same last two digits.
Current logic is hard coded to consider 0-69 to be in 2000-2069, and 70-99 to be 1970-1999, instead of comparing versus the current year.

...
\end{lstlisting}

\begin{lstlisting}[breaklines=true, basicstyle=\small\ttfamily]
<@content@>:
...
### django/utils/http.py
...

    `epoch_seconds` is a floating point number expressed in seconds since the
    epoch, in UTC - such as that outputted by time.time(). If set to None, it
    defaults to the current time.

    Output a string in the format 'Wdy, DD Mon YYYY HH:MM:SS GMT'.
    """
    return formatdate(epoch_seconds, usegmt=True)


def parse_http_date(date):
    """
    Parse a date format as specified by HTTP RFC7231 section 7.1.1.1.

    The three formats allowed by the RFC are accepted, even if only the first
    one is still in widespread use.

    Return an integer expressed in seconds since the epoch, in UTC.
    """
    # email.utils.parsedate() does the job for RFC1123 dates; unfortunately
    # RFC7231 makes it mandatory to support RFC850 dates too. So we roll
    # our own RFC-compliant parsing.
    for regex in RFC1123_DATE, RFC850_DATE, ASCTIME_DATE:
        m = regex.match(date)
        if m is not None:
            break
    else:
        raise ValueError("%r is not in a valid HTTP date format" % date)
    try:
        year = int(m.group('year'))
        if year < 100:
            if year < 70:
                year += 2000
            else:
                year += 1900
        month = MONTHS.index(m.group('mon').lower()) + 1
        day = int(m.group('day'))
        hour = int(m.group('hour'))
        min = int(m.group('min'))
        sec = int(m.group('sec'))
        result = datetime.datetime(year, month, day, hour, min, sec)
        return calendar.timegm(result.utctimetuple())
    except Exception as exc:
        raise ValueError("%r is not a valid date" % date) from exc


def parse_http_date_safe(date):
    """
    Same as parse_http_date, but return None if the input is invalid.
    """
    try:
        return parse_http_date(date)
    except Exception:
        pass
...
\end{lstlisting}

\newpage
\subsection{File-level Granularity Content Example}

The \texttt{\{problem\_statement\}} and \texttt{\{content\}} placeholders at file-level granularity for instance \lstinline{django__django-11848} are shown below.

\begin{lstlisting}[breaklines=true, basicstyle=\small\ttfamily]
<@problem_statement@>:

django.utils.http.parse_http_date two digit year check is incorrect
Description
	 
		(last modified by Ad Timmering)
	 
RFC 850 does not mention this, but in RFC 7231 (and there's something similar in RFC 2822), there's the following quote:
Recipients of a timestamp value in rfc850-date format, which uses a
two-digit year, MUST interpret a timestamp that appears to be more
than 50 years in the future as representing the most recent year in
the past that had the same last two digits.
Current logic is hard coded to consider 0-69 to be in 2000-2069, and 70-99 to be 1970-1999, instead of comparing versus the current year.

\end{lstlisting}

\begin{lstlisting}[breaklines=true, basicstyle=\small\ttfamily]
<@content@>:

### django/utils/http.py
...
import base64
import calendar
import datetime
import re
import unicodedata
import warnings
from binascii import Error as BinasciiError
from email.utils import formatdate
from urllib.parse import (
    ParseResult, SplitResult, _coerce_args, _splitnetloc, _splitparams, quote,
    quote_plus, scheme_chars, unquote, unquote_plus,
    urlencode as original_urlencode, uses_params,
)

from django.core.exceptions import TooManyFieldsSent
from django.utils.datastructures import MultiValueDict
from django.utils.deprecation import RemovedInDjango40Warning
from django.utils.functional import keep_lazy_text

# based on RFC 7232, Appendix C
ETAG_MATCH = re.compile(r'''
    \A(      # start of string and capture group
    (?:W/)?  # optional weak indicator
    "        # opening quote
    [^"]*    # any sequence of non-quote characters
    "        # end quote
    )\Z      # end of string and capture group
''', re.X)

MONTHS = 'jan feb mar apr may jun jul aug sep oct nov dec'.split()
__D = r'(?P<day>\d{2})'
__D2 = r'(?P<day>[ \d]\d)'
__M = r'(?P<mon>\w{3})'
__Y = r'(?P<year>\d{4})'
__Y2 = r'(?P<year>\d{2})'
__T = r'(?P<hour>\d{2}):(?P<min>\d{2}):(?P<sec>\d{2})'
RFC1123_DATE = re.compile(r'^\w{3}, %s %s %s %s GMT$' % (__D, __M, __Y, __T))
RFC850_DATE = re.compile(r'^\w{6,9}, %s-%s-%s %s GMT$' % (__D, __M, __Y2, __T))
ASCTIME_DATE = re.compile(r'^\w{3} %s %s %s %s$' % (__M, __D2, __T, __Y))

RFC3986_GENDELIMS = ":/?#[]@"
RFC3986_SUBDELIMS = "!$&'()*+,;="

FIELDS_MATCH = re.compile('[&;]')


@keep_lazy_text
def urlquote(url, safe='/'):
    """
    A legacy compatibility wrapper to Python's urllib.parse.quote() function.
    (was used for unicode handling on Python 2)
    """
    warnings.warn(
        'django.utils.http.urlquote() is deprecated in favor of '
        'urllib.parse.quote().',
        RemovedInDjango40Warning, stacklevel=2,
    )
    return quote(url, safe)


@keep_lazy_text
def urlquote_plus(url, safe=''):
    """
    A legacy compatibility wrapper to Python's urllib.parse.quote_plus()
    function. (was used for unicode handling on Python 2)
    """
    warnings.warn(
        'django.utils.http.urlquote_plus() is deprecated in favor of '
        'urllib.parse.quote_plus(),',
        RemovedInDjango40Warning, stacklevel=2,
    )
    return quote_plus(url, safe)


@keep_lazy_text
def urlunquote(quoted_url):
    """
    A legacy compatibility wrapper to Python's urllib.parse.unquote() function.
    (was used for unicode handling on Python 2)
    """
    warnings.warn(
        'django.utils.http.urlunquote() is deprecated in favor of '
        'urllib.parse.unquote().',
        RemovedInDjango40Warning, stacklevel=2,
    )
    return unquote(quoted_url)


@keep_lazy_text
def urlunquote_plus(quoted_url):
    """
    A legacy compatibility wrapper to Python's urllib.parse.unquote_plus()
    function. (was used for unicode handling on Python 2)
    """
    warnings.warn(
        'django.utils.http.urlunquote_plus() is deprecated in favor of '
        'urllib.parse.unquote_plus().',
        RemovedInDjango40Warning, stacklevel=2,
    )
    return unquote_plus(quoted_url)


def urlencode(query, doseq=False):
    """
    A version of Python's urllib.parse.urlencode() function that can operate on
    MultiValueDict and non-string values.
    """
    if isinstance(query, MultiValueDict):
        query = query.lists()
    elif hasattr(query, 'items'):
        query = query.items()
    query_params = []
    for key, value in query:
        if value is None:
            raise TypeError(
                "Cannot encode None for key '%s' in a query string. Did you "
                "mean to pass an empty string or omit the value?" % key
            )
        elif not doseq or isinstance(value, (str, bytes)):
            query_val = value
        else:
            try:
                itr = iter(value)
            except TypeError:
                query_val = value
            else:
                # Consume generators and iterators, when doseq=True, to
                # work around https://bugs.python.org/issue31706.
                query_val = []
                for item in itr:
                    if item is None:
                        raise TypeError(
                            "Cannot encode None for key '%s' in a query "
                            "string. Did you mean to pass an empty string or "
                            "omit the value?" % key
                        )
                    elif not isinstance(item, bytes):
                        item = str(item)
                    query_val.append(item)
        query_params.append((key, query_val))
    return original_urlencode(query_params, doseq)


def http_date(epoch_seconds=None):
    """
    Format the time to match the RFC1123 date format as specified by HTTP
    RFC7231 section 7.1.1.1.

    `epoch_seconds` is a floating point number expressed in seconds since the
    epoch, in UTC - such as that outputted by time.time(). If set to None, it
    defaults to the current time.

    Output a string in the format 'Wdy, DD Mon YYYY HH:MM:SS GMT'.
    """
    return formatdate(epoch_seconds, usegmt=True)


def parse_http_date(date):
    """
    Parse a date format as specified by HTTP RFC7231 section 7.1.1.1.

    The three formats allowed by the RFC are accepted, even if only the first
    one is still in widespread use.

    Return an integer expressed in seconds since the epoch, in UTC.
    """
    # email.utils.parsedate() does the job for RFC1123 dates; unfortunately
    # RFC7231 makes it mandatory to support RFC850 dates too. So we roll
    # our own RFC-compliant parsing.
    for regex in RFC1123_DATE, RFC850_DATE, ASCTIME_DATE:
        m = regex.match(date)
        if m is not None:
            break
    else:
        raise ValueError("%r is not in a valid HTTP date format" % date)
    try:
        year = int(m.group('year'))
        if year < 100:
            if year < 70:
                year += 2000
            else:
                year += 1900
        month = MONTHS.index(m.group('mon').lower()) + 1
        day = int(m.group('day'))
        hour = int(m.group('hour'))
        min = int(m.group('min'))
        sec = int(m.group('sec'))
        result = datetime.datetime(year, month, day, hour, min, sec)
        return calendar.timegm(result.utctimetuple())
    except Exception as exc:
        raise ValueError("%r is not a valid date" % date) from exc


def parse_http_date_safe(date):
    """
    Same as parse_http_date, but return None if the input is invalid.
    """
    try:
        return parse_http_date(date)
    except Exception:
        pass


# Base 36 functions: useful for generating compact URLs

def base36_to_int(s):
    """
    Convert a base 36 string to an int. Raise ValueError if the input won't fit
    into an int.
    """
    # To prevent overconsumption of server resources, reject any
    # base36 string that is longer than 13 base36 digits (13 digits
    # is sufficient to base36-encode any 64-bit integer)
    if len(s) > 13:
        raise ValueError("Base36 input too large")
    return int(s, 36)


def int_to_base36(i):
    """Convert an integer to a base36 string."""
    char_set = '0123456789abcdefghijklmnopqrstuvwxyz'
    if i < 0:
        raise ValueError("Negative base36 conversion input.")
    if i < 36:
        return char_set[i]
    b36 = ''
    while i != 0:
        i, n = divmod(i, 36)
        b36 = char_set[n] + b36
    return b36


def urlsafe_base64_encode(s):
    """
    Encode a bytestring to a base64 string for use in URLs. Strip any trailing
    equal signs.
    """
    return base64.urlsafe_b64encode(s).rstrip(b'\n=').decode('ascii')


def urlsafe_base64_decode(s):
    """
    Decode a base64 encoded string. Add back any trailing equal signs that
    might have been stripped.
    """
    s = s.encode()
    try:
        return base64.urlsafe_b64decode(s.ljust(len(s) + len(s) % 4, b'='))
    except (LookupError, BinasciiError) as e:
        raise ValueError(e)


def parse_etags(etag_str):
    """
    Parse a string of ETags given in an If-None-Match or If-Match header as
    defined by RFC 7232. Return a list of quoted ETags, or ['*'] if all ETags
    should be matched.
    """
    if etag_str.strip() == '*':
        return ['*']
    else:
        # Parse each ETag individually, and return any that are valid.
        etag_matches = (ETAG_MATCH.match(etag.strip()) for etag in etag_str.split(','))
        return [match.group(1) for match in etag_matches if match]


def quote_etag(etag_str):
    """
    If the provided string is already a quoted ETag, return it. Otherwise, wrap
    the string in quotes, making it a strong ETag.
    """
    if ETAG_MATCH.match(etag_str):
        return etag_str
    else:
        return '"%s"' % etag_str


def is_same_domain(host, pattern):
    """
    Return ``True`` if the host is either an exact match or a match
    to the wildcard pattern.

    Any pattern beginning with a period matches a domain and all of its
    subdomains. (e.g. ``.example.com`` matches ``example.com`` and
    ``foo.example.com``). Anything else is an exact string match.
    """
    if not pattern:
        return False

    pattern = pattern.lower()
    return (
        pattern[0] == '.' and (host.endswith(pattern) or host == pattern[1:]) or
        pattern == host
    )


def url_has_allowed_host_and_scheme(url, allowed_hosts, require_https=False):
    """
    Return ``True`` if the url uses an allowed host and a safe scheme.

    Always return ``False`` on an empty url.

    If ``require_https`` is ``True``, only 'https' will be considered a valid
    scheme, as opposed to 'http' and 'https' with the default, ``False``.

    Note: "True" doesn't entail that a URL is "safe". It may still be e.g.
    quoted incorrectly. Ensure to also use django.utils.encoding.iri_to_uri()
    on the path component of untrusted URLs.
    """
    if url is not None:
        url = url.strip()
    if not url:
        return False
    if allowed_hosts is None:
        allowed_hosts = set()
    elif isinstance(allowed_hosts, str):
        allowed_hosts = {allowed_hosts}
    # Chrome treats \ completely as / in paths but it could be part of some
    # basic auth credentials so we need to check both URLs.
    return (
        _url_has_allowed_host_and_scheme(url, allowed_hosts, require_https=require_https) and
        _url_has_allowed_host_and_scheme(url.replace('\\', '/'), allowed_hosts, require_https=require_https)
    )


def is_safe_url(url, allowed_hosts, require_https=False):
    warnings.warn(
        'django.utils.http.is_safe_url() is deprecated in favor of '
        'url_has_allowed_host_and_scheme().',
        RemovedInDjango40Warning, stacklevel=2,
    )
    return url_has_allowed_host_and_scheme(url, allowed_hosts, require_https)


# Copied from urllib.parse.urlparse() but uses fixed urlsplit() function.
def _urlparse(url, scheme='', allow_fragments=True):
    """Parse a URL into 6 components:
    <scheme>://<netloc>/<path>;<params>?<query>#<fragment>
    Return a 6-tuple: (scheme, netloc, path, params, query, fragment).
    Note that we don't break the components up in smaller bits
    (e.g. netloc is a single string) and we don't expand % escapes."""
    url, scheme, _coerce_result = _coerce_args(url, scheme)
    splitresult = _urlsplit(url, scheme, allow_fragments)
    scheme, netloc, url, query, fragment = splitresult
    if scheme in uses_params and ';' in url:
        url, params = _splitparams(url)
    else:
        params = ''
    result = ParseResult(scheme, netloc, url, params, query, fragment)
    return _coerce_result(result)


# Copied from urllib.parse.urlsplit() with
# https://github.com/python/cpython/pull/661 applied.
def _urlsplit(url, scheme='', allow_fragments=True):
    """Parse a URL into 5 components:
    <scheme>://<netloc>/<path>?<query>#<fragment>
    Return a 5-tuple: (scheme, netloc, path, query, fragment).
    Note that we don't break the components up in smaller bits
    (e.g. netloc is a single string) and we don't expand % escapes."""
    url, scheme, _coerce_result = _coerce_args(url, scheme)
    netloc = query = fragment = ''
    i = url.find(':')
    if i > 0:
        for c in url[:i]:
            if c not in scheme_chars:
                break
        else:
            scheme, url = url[:i].lower(), url[i + 1:]

    if url[:2] == '//':
        netloc, url = _splitnetloc(url, 2)
        if (('[' in netloc and ']' not in netloc) or
                (']' in netloc and '[' not in netloc)):
            raise ValueError("Invalid IPv6 URL")
    if allow_fragments and '#' in url:
        url, fragment = url.split('#', 1)
    if '?' in url:
        url, query = url.split('?', 1)
    v = SplitResult(scheme, netloc, url, query, fragment)
    return _coerce_result(v)


def _url_has_allowed_host_and_scheme(url, allowed_hosts, require_https=False):
    # Chrome considers any URL with more than two slashes to be absolute, but
    # urlparse is not so flexible. Treat any url with three slashes as unsafe.
    if url.startswith('///'):
        return False
    try:
        url_info = _urlparse(url)
    except ValueError:  # e.g. invalid IPv6 addresses
        return False
    # Forbid URLs like http:///example.com - with a scheme, but without a hostname.
    # In that URL, example.com is not the hostname but, a path component. However,
    # Chrome will still consider example.com to be the hostname, so we must not
    # allow this syntax.
    if not url_info.netloc and url_info.scheme:
        return False
    # Forbid URLs that start with control characters. Some browsers (like
    # Chrome) ignore quite a few control characters at the start of a
    # URL and might consider the URL as scheme relative.
    if unicodedata.category(url[0])[0] == 'C':
        return False
    scheme = url_info.scheme
    # Consider URLs without a scheme (e.g. //example.com/p) to be http.
    if not url_info.scheme and url_info.netloc:
        scheme = 'http'
    valid_schemes = ['https'] if require_https else ['http', 'https']
    return ((not url_info.netloc or url_info.netloc in allowed_hosts) and
            (not scheme or scheme in valid_schemes))


def limited_parse_qsl(qs, keep_blank_values=False, encoding='utf-8',
                      errors='replace', fields_limit=None):
    """
    Return a list of key/value tuples parsed from query string.

    Copied from urlparse with an additional "fields_limit" argument.
    Copyright (C) 2013 Python Software Foundation (see LICENSE.python).

    Arguments:

    qs: percent-encoded query string to be parsed

    keep_blank_values: flag indicating whether blank values in
        percent-encoded queries should be treated as blank strings. A
        true value indicates that blanks should be retained as blank
        strings. The default false value indicates that blank values
        are to be ignored and treated as if they were  not included.

    encoding and errors: specify how to decode percent-encoded sequences
        into Unicode characters, as accepted by the bytes.decode() method.

    fields_limit: maximum number of fields parsed or an exception
        is raised. None means no limit and is the default.
    """
    if fields_limit:
        pairs = FIELDS_MATCH.split(qs, fields_limit)
        if len(pairs) > fields_limit:
            raise TooManyFieldsSent(
                'The number of GET/POST parameters exceeded '
                'settings.DATA_UPLOAD_MAX_NUMBER_FIELDS.'
            )
    else:
        pairs = FIELDS_MATCH.split(qs)
    r = []
    for name_value in pairs:
        if not name_value:
            continue
        nv = name_value.split('=', 1)
        if len(nv) != 2:
            # Handle case of a control-name with no equal sign
            if keep_blank_values:
                nv.append('')
            else:
                continue
        if nv[1] or keep_blank_values:
            name = nv[0].replace('+', ' ')
            name = unquote(name, encoding=encoding, errors=errors)
            value = nv[1].replace('+', ' ')
            value = unquote(value, encoding=encoding, errors=errors)
            r.append((name, value))
    return r


def escape_leading_slashes(url):
    """
    If redirecting to an absolute path (two leading slashes), a slash must be
    escaped to prevent browsers from handling the path as schemaless and
    redirecting to another host.
    """
    if url.startswith('//'):
        url = '/%2F{}'.format(url[2:])
    return url
...
\end{lstlisting}

\subsection{Model Configuration}

Repair experiments were conducted using a vLLM-served model with the following configuration:

\vspace{0.5em}
\noindent
\begin{tabular}{@{}ll@{}}
\toprule
\textbf{Parameter} & \textbf{Value} \\
\midrule
\texttt{--model} & \texttt{Qwen/Qwen3-Coder-30B-A3B-Instruct} \\
\texttt{--tensor-parallel-size} & \texttt{1} \\
\texttt{--gpu-memory-utilization} & \texttt{0.95} \\
\texttt{--max-num-seqs} & \texttt{32} \\
\texttt{--max-model-len} & \texttt{262144} \\
\texttt{--enable-chunked-prefill} & \texttt{True} \\
\texttt{--enable-prefix-caching} & \texttt{True} \\
\texttt{--tool-call-parser} & \texttt{qwen3\_coder} \\
\texttt{--enable-auto-tool-choice} & \texttt{True} \\
\texttt{--dtype} & \texttt{bfloat16} \\
\texttt{--port} & \texttt{8000} \\
\bottomrule
\end{tabular}

\end{document}